# DESIGN OF A LOW-POWER 1.65 GBPS DATA CHANNEL FOR HDMI TRANSMITTER

Ajay Agrawal and R.S.Gamad

Department of Electronics and Instrumentation Engineering,
Shri Govindram Seksaria Institute of Technology and Science,
23, Park Road, Indore, M.P. 452003, India

*ABSTRACT*

*This paper presents a design of low power data channel for application in High Definition Multimedia Interface (HDMI) Transmitter circuit. The input is 10 bit parallel data and output is serial data at 1.65 Gbps. This circuit uses only a single frequency of serial clock input. All other timing signals are derived within the circuit from the serial clock. This design has dedicated lines to disable and enable all its channels within two pixel-clock periods only. A pair of disable and enable functions performed immediately after power-on of the circuit serves as the reset function. The presented design is immune to data-dependent switching spikes in supply current and pushes them in the range of serial frequency and its multiples. Thus filtering requirements are relaxed. The output stage uses a bias voltage of 2.8 volts for a receiver pull-up voltage of 3.3 volts. The reported data channel is designed using UMC 180 nm CMOS Technology. The design is modifiable for other inter-board serial interfaces like USB and LAN with different number of bits at the parallel input.*

## 1. INTRODUCTION

Presently, a large number of computer devices under development are hand-held mobile devices equipped with high speed serial communication ports like USB and HDMI. As these devices are battery operated, it is highly necessary that the power consumption of their transmitting circuits is kept as low as possible. Since the speed of transmission cannot be reduced for a standard real time data transmitter like HDMI, it is required to design the transmitter circuit without sacrificing the speed of transmission [1]. Some design use the feature of power harvesting from the receiver signalling channels allowing the high power consumption of the transmitter IC. Another solution is technology scaling for low power achievement at the desired speed. This normally results in increased design and fabrication cost of the devices. While some other designs use reduced signal swing of the output voltage which do not apply to the presented target interface [2-6]. This design presents the data channel section of the HDMI transmitter circuit in 180 nm CMOS which is input with only one frequency of serial clock coming from the PLL and eliminated the use of multiple clock signals [7]. It is powered with the standard 1.8V supply and driver cascade stage used a bias voltage of 2.8 V which is less than battery voltage of almost all mobile devices.

This paper is organized as follows. In section two, the overview of proposed HDMI transmitter data channel is discussed. Section three presents schematic of proposed data channel. The simulation results of proposed design are presented in section four and concluded in section five.

## 2. HDMI TRANSMITTER DATA CHANNEL

The block diagram of proposed HDMI data channel is shown in Fig.1. The transmitter data channel was fed with 10-bit parallel data from the video data controller. The FIFO and serializing block do the function of holding the 10-bit data and its serialization.

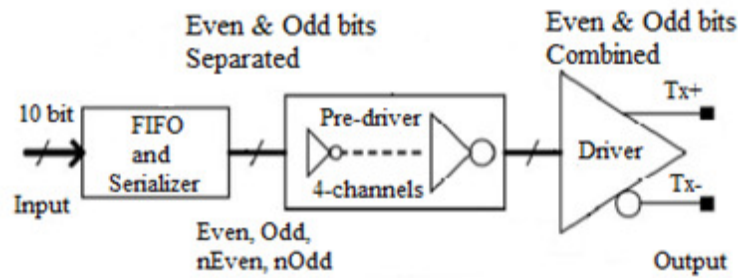

Figure 1. Block Diagram of Data Channel

The even 5 bits were multiplexed into one active low 'Even' signal and odd 5 bits were multiplexed into one active low 'Odd' signal. Similarly active low even and odd complements were also generated. The two 'Even' and 'Odd' signals were sufficiently buffered by two different pre-drivers with pull-up at their inputs. Two more similar pre-drivers were required for their complements. Thus four pre-driver channels were used. This even and odd period driving has the advantage of hiding data-dependent surges on supply nets which arise due to pre-driver switching [2]. The pre-driver outputs were applied to the driving circuit as shown in Fig. 1 to drive the output differential line. The driver output pins Tx+ and Tx- sink the current supplied from the receiver supply via termination resistors on the remote side of the HDMI cable. The receiver supply voltage is generally 3.3 Volts. Also as per the specifications[1] when the transmitter is in standby mode, the voltage drop across the resistors should not exceed 10mV each.

## 3. PROPOSED DATA CHANNEL

The schematic of presented Data Channel is shown in Fig. 2. It contains a tandem connected chain of ten falling-edge triggered clocked D-type flip-flops[8]. This flip-flop chain is shown in the lower part of Fig. 2 with outputs labelled *Sel1* to *Sel10*. In Fig. 2, the first flip-flop has a parallel similar flip-flop (left-most) and both their inputs are fed with the common signal *Start*. The output of first flip-flop is *Sel1* and that of its parallel is *iSel1*. These 11 flip-flops are triggered with two nearly complement clock signals *Dclk* and *Nclk* obtained from the input *Clock* signal by the SPLITTER circuit shown in Fig. 3.

A RESET circuit was provided for disabling and enabling the data channel. The schematic of this RESET circuit is shown in Fig. 4. The signal *Start* is driven as high at falling edge of *Dclk* only if *Disable* input was low and *Enable* input was high at the previous rising edge of *Dclk*. The Enable is made low before the next rising edge of *Dclk*. Another cause of signal *Start* going high when *Disable* and *Enable* are both low but *Buffered_Sel10* (due to *Sel10*) goes high. If *Start* is high for just one clock period, then *Sel1* will be high in the next clock period and then *Sel2* and so on. The output propagates with every clock cycle up to *Sel10* and *Start* nearly coincides with *Sel10* via the FO4 BUFFER and RESET circuit. Now *Start* will cause *Sel1* to go high in the eleventh clock cycle. Thus a sequence of high for one input clock period is obtained at each output *Sel1* to *Sel10* one by one and the process keep repeating until *Disable* signal is raised. The data channel used ten selector circuits labelled as SEL in Fig. 2. Fig. 5 shows the schematic of SEL block. Each high output out-of *Sel1-Sel10* is used to select the corresponding bit of parallel input data *D0-D9* with the help of SEL blocks as shown in upper part of Fig. 2. The last two SEL blocks are not directly fed with *D8* and *D9*. Instead they are fed via two clocked D-type flip-flops. The first eight SEL blocks' and the last two bits' flip-flops respective outputs are updated for selection when *Sel9* goes high and cause suitable complement signals to be generated by the inverting and buffering circuit labelled FD. The two flip-flops were provided to retain correct (previously held) ninth and tenth bit for selection.

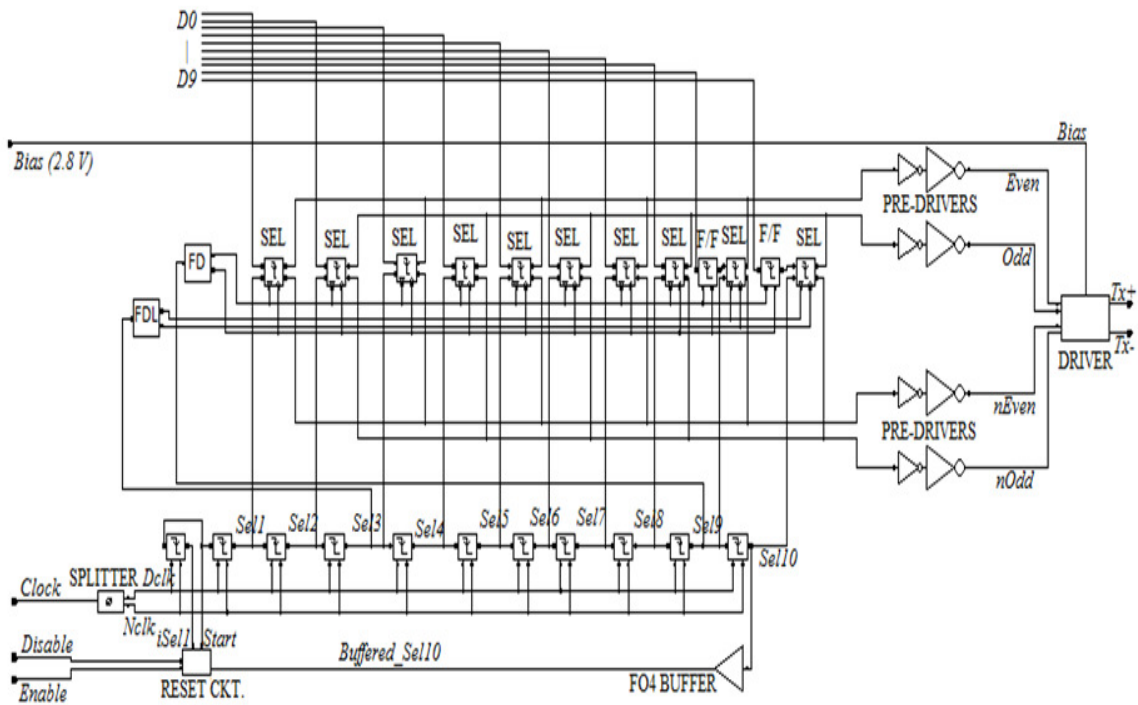

Figure 2. Schematic of proposed Data Channel

Figure 3.Schematic of clock splitter circuit

Figure 4. Schematic of reset circuit

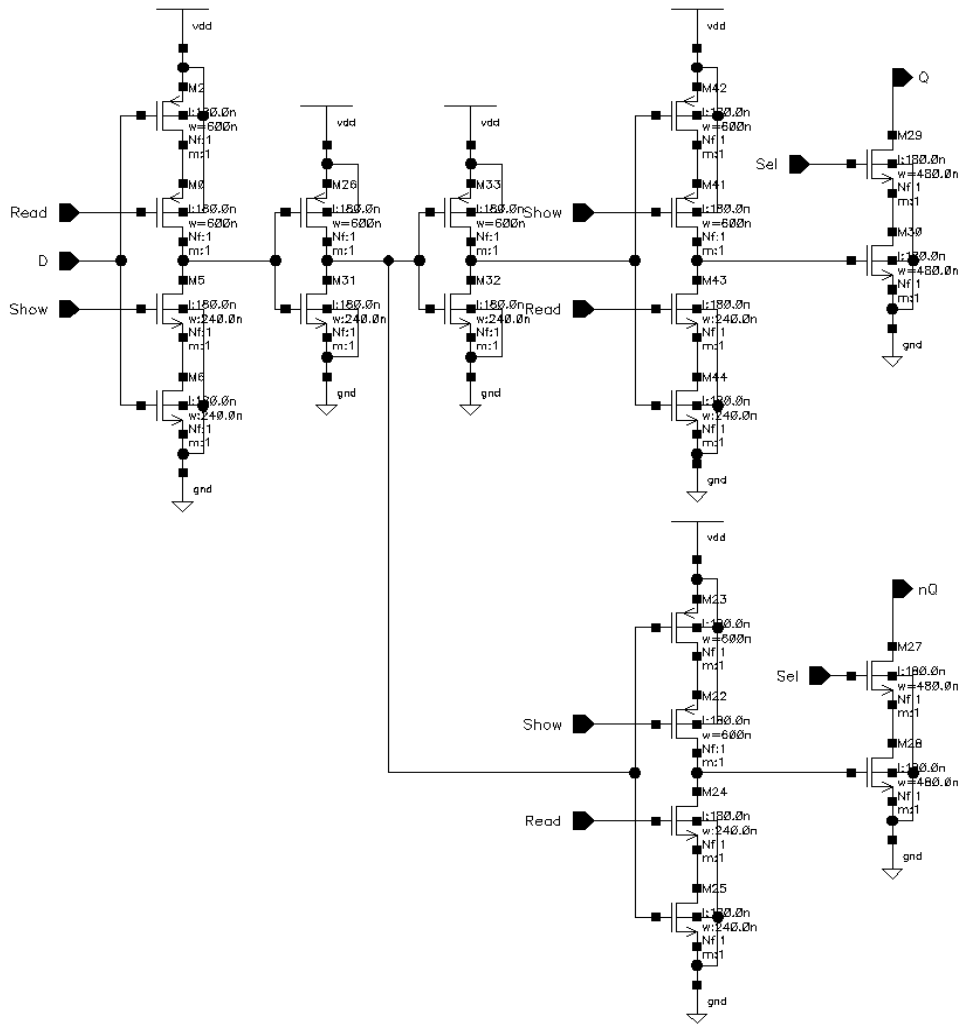

Figure 5. Schematic of selector block

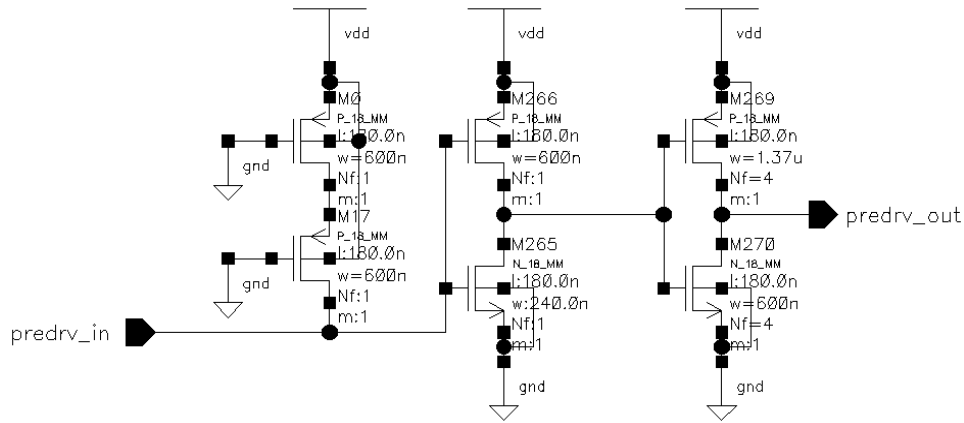

Figure 6. Schematic of pre-driver circuit

International Journal of VLSI design & Communication Systems (VLSICS) Vol.6, No.6, December 2015

These two flip-flops update the output with the falling and rising edge respective of their own clock (Show) and anti-clock (Read) inputs. These inputs are derived from *Sel3* high level by the inverting and buffering circuit labelled as FDL. Hence parallel 10 bit data is taken in hold before first bit selected and kept in hold until all bits are serialized. Thus all correct data bits are selected in one round of selection from *Sel1* to *Sel10*. This is the First-In First-Out (FIFO) action of the data channel.

The SELs' complementary outputs are connected in two groups of even and odd bit periods for the purpose of pseudo-shadow switching [9]. Thus four pre-drivers were used and each pre-driver has its own internal pull-up at the input as shown in Fig. 6 so that active low can be multiplexed from the five connected SEL outputs. The even and odd pre-driver outputs were used to drive the parallel pmos transistors in DRIVER circuit as shown in Fig.7. Thus 10:1 time-division multiplexing was achieved. A complementarily active-low driving pair was required for differential driving. Fig. 8 shows the proposed layout for the presented design.

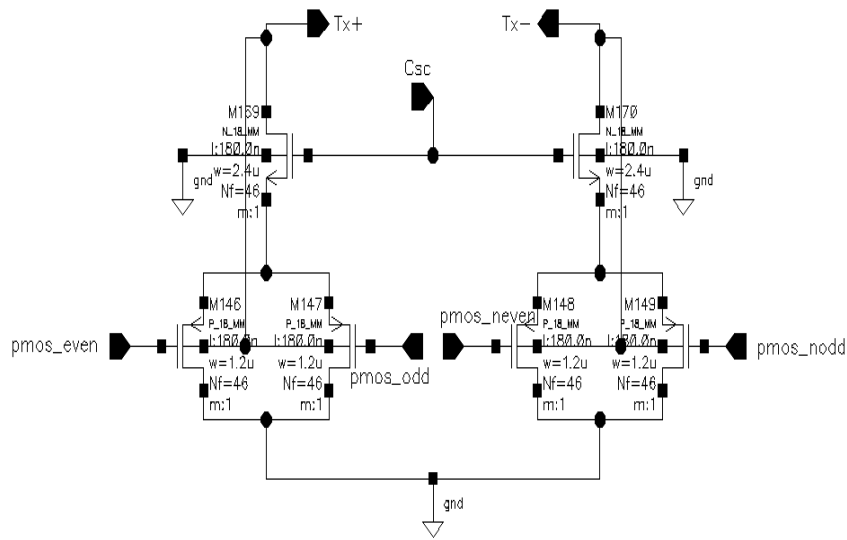

Figure 7. Schematic of driver circuit

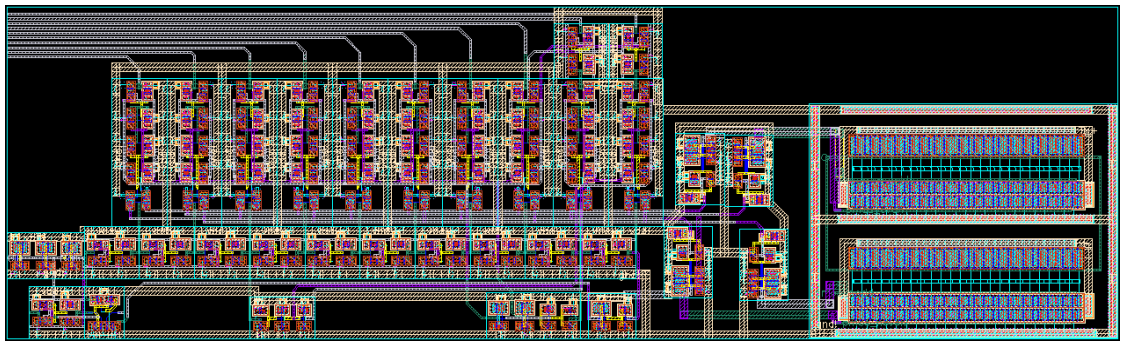

Figure 8. Proposed layout of Data Channel

## 4. SIMULATION RESULTS

The presented design was simulated using Cadence Virtuoso. During post-layout simulation the bit-select signals *Sel1-Sel10*, *iSel1* and *Start* signals were obtained as shown in Fig. 9. The results for schematic and post-layout simulation are compared in Table 1. The differential output eye diagram requirements of the HDMI specification and post-layout simulation are displayed in Fig. 10 and 11 respectively. Fig. 12 shows the spectrum of supply current drawn by three as such presented data channels and one clock channel. As shown in Fig. 12, supply current components within 500 MHz are less than 6% of dc component. The simulation results are compared against HDMI specification [1] for output voltage requirements in Table 2.

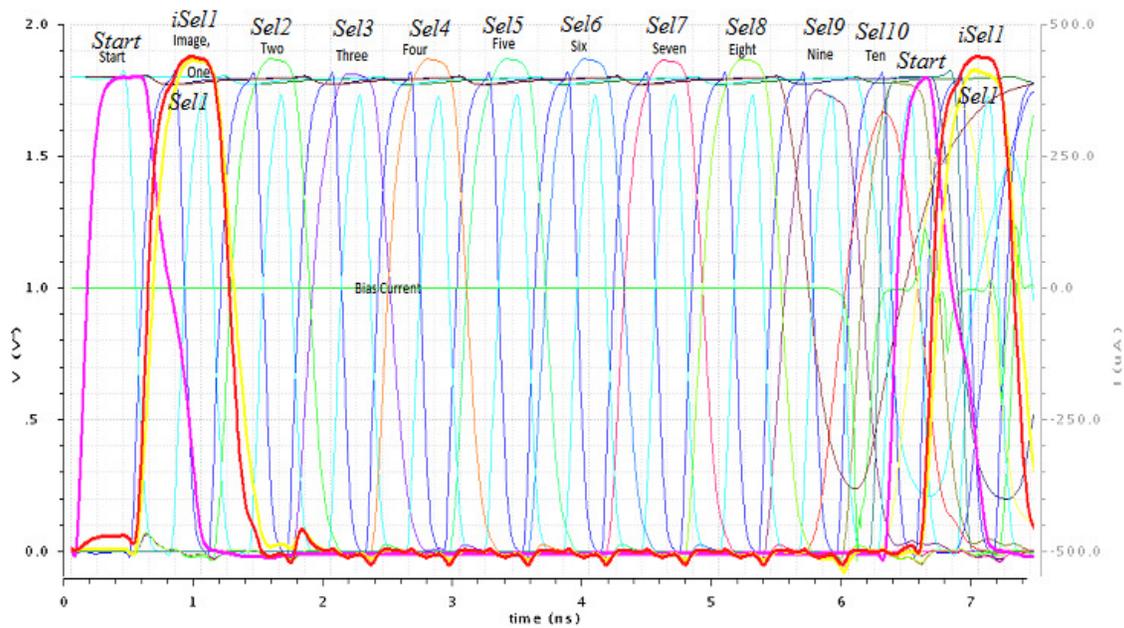

Figure 9. Bit-select and other signals

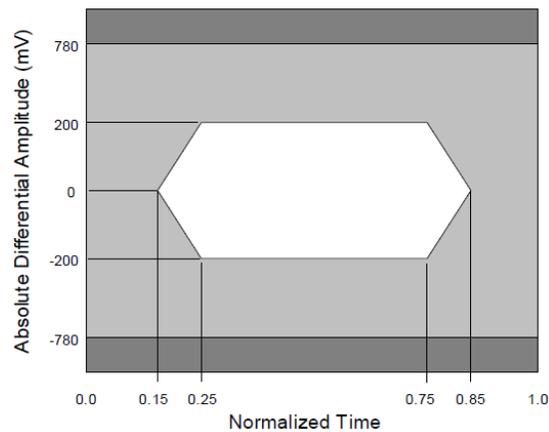

Figure 10. Eye diagram specification

29

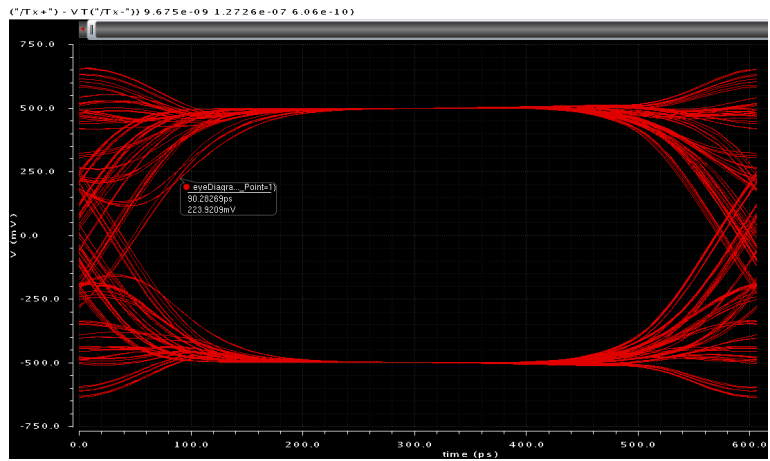

Figure 11. Post-layout simulation eye diagram

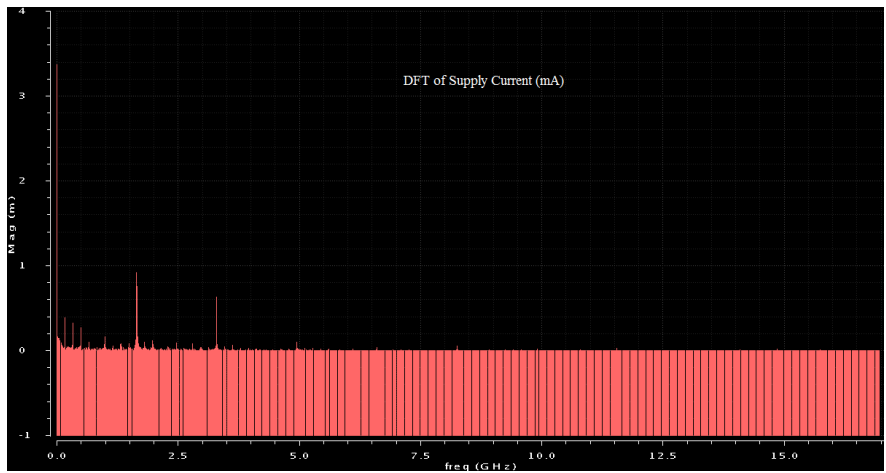

Figure 12. Spectrum of supply current

Table 1 Results of Data Channel Simulation

| Parameter | Schematic | Post-Layout |
|---|---|---|
| Supply Current at 1.8V | 678.4 uA | 1.005 mA |
| Single Ended High Voltage | 3.299 V | 3.299 V |
| Single Ended Low Voltage | 2.8019 V | 2.8019 V |
| Rise Time minimum | 108 ps | 104 ps |
| Fall Time minimum | 108 ps | 104 ps |
| Cascade Bias Current at 2.8 V | 597.2 nA | 534.3 nA |

Table 2. Performance Chart

| Item | Desired Value [1] | Achieved Value |
|---|---|---|
| Single-ended standby output voltage, $V_{off}$ | 3.290 V $\leq$ $V_{off}$ $\leq$ 3.310 V | 3.299 V |
| Single-ended output swing voltage, $V_{swing}$ | 400mV $\leq$ $V_{swing}$ $\leq$ 600mV | 660 mV |
| Single-ended high level output voltage, $V_H$ | 3.3 V $\pm$ 10mV | 3.299 V |
| Single-ended low level output voltage, $V_L$ | 2.7 V $\leq$ $V_L$ $\leq$ 2.9 V | 2.8019 V |
| Rise time & fall time (20%-80%) | 75 psec $\leq$ Rise time & 75 psec $\leq$ fall time | 104 psec & 104 psec |

## 5. CONCLUSIONS

In this paper, a design was presented for data channel of 1.65 Gbps serial speed for application in HDMI transmitter with low power consumption. The single ended output voltage swing could be manipulated by controlling the time of over-lap and under-lap of driver's parallel MOS transistors by providing additional combinational logic in the serializer. The presented schematic can also be utilized for other similar high speed serial interfaces like USB and LAN with applicable modifications.

### ACKNOWLEDGEMENTS

The authors would like to thank the Department of Electronics and Instrumentation Engineering, S.G.S.I.T.S., Indore, India for providing the laboratory facility for performing this work.

### REFERENCES

[1] HDMI Licensing, LLC, High-Definition Multimedia Interface Specification Version 1.3a, http://www.hdmi.org

[2] N. Gupta, T. Nandy, P. S. Sahni, M. Garg, J. N. Tripathi, (2014) "Zero Power 4.95Gbps HDMI Transmitter", IEEE International Symposium on Circuits and Systems (ISCAS), Melbourne VIC, page no. 1500 – 1503.

[3] N. Gupta, P. Bala, V.K. Singh, (2013) "Area and Power Efficient 3.4Gbps/Channel HDMI Transmitter with Single-Ended Structure", Proceedings of the 26th International Conference on VLSI Design and 12th International Conference on Embedded Systems (VLSID), Pune, ISSN: 1063-9667, page no. 142 – 146.

[4] N. Gupta, T. Nandy, S. Kundu, (2012) "HDMI transmitter in 32nM technology using 28Å MOS", IEEE International Symposium on Circuits and Systems (ISCAS), Seoul, ISSN: 0271-4302, page no. 1951 – 1954.

[5]  Y. Jeong, et al., (2012) "0.37mW/Gb/s low power SLVS transmitter for battery powered applications", IEEE International Symposium on Circuits and Systems (ISCAS), Seoul, ISSN: 0271-4302, page no. 1955 – 1958.

[6]  R. Inti, et al., (2011) "A highly digital 0.5 -to-4Gb/s 1.9mW/Gb/s serial-link transceiver using current-recycling in 90nm CMOS", IEEE International Solid-State Circuits Conference Digest of Technical Papers (ISSCC), San Francisco, CA, ISSN: 0193-6530, page no.152 – 154.

[7]  K. Kim, K. Jung, C. Park, W. Park, S. Lee and S. Cho, (2010) "A 3.4Gbps Transmitter for Multi-Serial Data Communication using pre-emphasis method", Proceedings of the 4th WSEAS international conference on Circuits, systems, signal and telecommunications, pp. 153-156.

[8]  Y. Suzuki , K. Odagawa and T. Abe, (1973) "Clocked CMOS calculator circuitry", IEEE J. Solid-State Circuits, vol. 8, pp.462 -469.

[9]  N. Gupta, T. Nandy, P. Bala, (2012) "Self- -Induced Supply Noise Reduction Technique in GBPS Rate Transmitters", Proceedings of the 25th International Conference on VLSI Design (VLSID), pp. 92, 95, 7-11.